\documentclass[aps,pre,superscriptaddress,twocolumn,amsmath,amssymb,showpacs]{revtex4}
\usepackage{graphicx}% Include figure files
\usepackage{dcolumn}% Align table columns on decimal point
\usepackage{bm}% bold math
\newcommand{\ignore}[1]{}
\usepackage{epsfig}
\usepackage{color}

\newcommand{\bx}{{\bf  x}}
\newcommand{\br}{{\bf  r}}
\newcommand{\bk}{{\bf  k}}

\begin{document}

\title{Spatial patterns in mesic savannas: the local facilitation limit 
and the role of demographic stochasticity.}

\author{Ricardo Mart\'inez-Garc\'ia}
\affiliation{IFISC, Instituto de F\'isica Interdisciplinar y Sistemas Complejos (CSIC-UIB),
 E-07122 Palma de Mallorca, Spain}

\author{Justin M. Calabrese}
\affiliation{Conservation Ecology Center, Smithsonian Conservation Biology Institute, National Zoological Park,
1500 Remount Rd., Front Royal, VA 22630, USA.} 

\author{Crist\'obal L\'opez} 
\affiliation{IFISC, Instituto de F\'isica Interdisciplinar y Sistemas Complejos (CSIC-UIB),
 E-07122 Palma de Mallorca, Spain}

\begin{abstract}
We propose a model equation for the dynamics of tree density in mesic savannas. 
It considers long-range competition among trees and the effect of fire acting as a local facilitation mechanism. Despite short-range
facilitation is taken to the local-range limit, 
the standard full spectrum of spatial structures 
obtained in general vegetation models is recovered.
Long-range competition is thus the key ingredient for the development of patterns.
The long time coexistence between trees and grass, and how fires affect
the survival of trees as well as the maintenance of the patterns is studied. 
The influence of demographic noise is analyzed. The stochastic system,
under the parameter constraints typical of mesic savannas, shows irregular
patterns characteristics of realistic situations.  
The coexistence of trees and grass still remains at reasonable noise
intensities.
\end{abstract}

\maketitle

\section{Introduction}

Savannas ecosystems are characterized by the long-term coexistence between a continuous grass
layer and scattered or clustered trees \cite{sarmiento1984}. Occurring in many regions of the
world, in areas with very different climatic and ecological conditions, the spatial structure,
persistence, and resilience of savannas have long intrigued ecologists \cite{scholes, sankaran2005, borgogno,belsky1994influences}.
However, despite substantial research, the origin and nature of savannas have not yet been fully resolved and much remains to be learned. 

Savanna tree populations often exhibit pronounced, non-random spatial
structures \cite{skarpe1991spatial, barot1999demography, jeltsch1999detecting, caylor2003tree, scanlon2007positive}.
Much research has therefore focused on explaining how spatial patterning in savannas arises
\cite{jeltsch,jeltsch1999detecting,scanlon2007positive,skarpe1991spatial,justin,vazquez}. In most natural plant systems both facilitative and competitive processes are simultaneously
present \cite{scholes,vetaas} and hard to disentangle \cite{veblen,barbier}. Some savanna studies have pointed toward the existence of 
short-distance facilitation \cite{caylor2003tree, scanlon2007positive}, while others have demonstrated evidence of competition
\cite{skarpe1991spatial, jeltsch1999detecting, barot1999demography}, with conflicting reports sometimes arriving from the same regions.

Different classes of savannas, which can be characterized by how much rainfall they typically receive, should be affected 
by different sets of processes. For example, in semiarid savannas water is extremely limited (low mean annual precipitation)
and competition among trees is expected to be strong, but fire plays little role because there is typically not enough grass
biomass to serve as fuel. In contrast, humid savannas should be characterized by weaker competition among trees, but also by
frequent and intense fires. In-between these extremes, in mesic savannas, trees likely have to contend with intermediate levels
of both competition for water and fire \cite{justin,sankaran2005, sankaran2008, bond2003, bond2008, bucini}. 

Competition among trees is mediated by roots that typically extend well beyond the crown \cite{borgogno,barbier}. Additionally, fire
can lead to local facilitation due to a protection effect, whereby vulnerable juvenile trees placed near adults are
protected from fire by them \cite{holdo}. We are particularly interested in how the interplay between these mechanisms
governs the spatial arrangement of trees in mesic savannas, where both mechanisms may operate. 
On the other side, it has frequently been claimed
that pattern formation in arid systems can be explained by a combination of long-distance competition and short-distance facilitation 
\cite{klausmeier, Lefever, lefeverJTB2009, Lefever2012, rietkirkAmNat2002, vonHardenberg,dodorico}. This combination
of mechanisms is also known to produce spatial structures in many other natural systems \cite{cross}. Although mesic
savannas do not display the same range of highly regular spatial patterns that arise in arid systems (e.g., tigerbush),
similar mechanisms might be at work. Specifically, the interaction between long-range competition and short-range facilitation
might still play a role in pattern formation in savanna tree populations, 
but only for a limited range of parameter values and possibly modified by demographic stochasticity.

Although the facilitation component has often been thought to be a key component in previous vegetation
models \cite{dodorico, dodorico2006b, rietkirkAmNat2002, scanlon2007positive}, Rietkerk and Van de Koppel \cite{rietkirkRev2008}, 
speculated, but did not show, that pattern formation could occur without short-range facilitation in the particular example of tidal freshwater marsh.
In the case of savannas, as stated before, the presence of adult trees favor the establishment of new trees in the area, protecting the juveniles against fires. Considering this effect, 
we take the facilitation component to its infinitesimally short spatial limit, and study its effect in the emergence of spatially periodic structures of trees.
To our knowledge, this explanation, and the interrelation between long-range competition and local facilitation,
has not been explored for a vegetation system. One of our main results is that when considering the limit of local facilitation and nonlocal competition, clustering of trees appears.

Here we develop a minimalistic model of savannas that considers two of
the factors, as already mentioned,  
thought to be crucial to structure
mesic savannas: tree-tree competition and fire, with a primary focus on spatially nonlocal competition. 
Employing standard tools used in the study of pattern formation phenomena
in physics (stability analysis and the structure function) \cite{cross}, we explore
the conditions under which the model can produce non-homogeneous spatial distributions.
A key strength of our approach is that we are able to provide a complete and rigorous analysis of 
the patterns the model is capable of producing, and we identify which among these correspond to situations that are relevant for mesic savannas.
We further examine
the role of demographic stochasticity in modifying both spatial patterns and the conditions under which trees persist in the system
in the presence of fire, and discuss the implications of these results for the debate on whether the balance of processes affecting savanna
trees is positive, negative, or is variable among systems. 
This is the framework of our study: the role of long-range competition, facilitation and demographic fluctuations (in the second
part of the paper) in the spatial structures of mesic savannas.
To complete our work we include an 
appendix where we study the effect of external fluctuations (mimicking for e.g. rainfall) on
savanna dynamics.

Our model is inspired by the one presented by Calabrese et al. in
\cite{justin}. It complements theirs by providing further analytical results
that clearly demonstrate that this simple system, where we focus on the local limit of facilitation, can
produce the full spectrum of spatial patterns reported from models employing both short-range
facilitation and long-range inhibition (competition).

\section{The Deterministic Model} \label{deterministic}

In this section we derive the deterministic equation for the local density of trees,
such that dynamics is of the logistic type and we only consider
tree-tree competition and fire. We study the formation of patterns via stability analysis and provide numerical simulations 
of our model, showing the emergence of spatial structures.

\subsection{The nonlocal savanna model}\label{savannamodel}

Calabrese et al. \cite{justin} introduced a simple discrete-particle lattice savanna model that considers the birth-death
dynamics of trees, and where tree-tree competition and fire
are the principal ingredients. These mechanisms act on the probability of
establishment of a tree once a seed lands at a particular point on the lattice. In the discrete model, 
seeds land in the neighborhood of a parent tree with a rate $b$, and establish as adult trees if they are able to survive both competition neighboring trees and fire.
 As these two phenomena are independent, the probability of establishment is $P_{E}=P_{C}P_{F}$, where
$P_{C}$ is the probability of surviving the competition, and $P_{F}$ is the probability of surviving a fire event.
From this dynamics, we write a 
deterministic differential equation describing
the time evolution of the global density of trees (mean field), $\rho (t)$, where the population
has logistic growth at rate $b$, and an exponential death term at rate $\alpha$.
It reads:
\begin{equation}
\frac{d\rho}{dt}=bP_{E}(\rho) \rho(t)\left( 1-\rho(t)\right) -\alpha\rho(t).
\label{eq:mf}
\end{equation}

Generalizing Eq.~(\ref{eq:mf}), we propose an evolution equation for the space-dependent (local) density of trees,
$\rho (\bx, t)$:
\begin{equation}
\frac{\partial \rho(\bx, t)}{\partial t}=bP_{E}\rho(\bx, t)(1-\rho(\bx, t))-\alpha\rho(\bx, t).
\label{eq:PDE}
\end{equation}
We allow the probability of overcoming competition to 
depend on tree crowding in a local neighborhood, decaying exponentially with the
density of surrounding trees as
\begin{equation}
 P_{C}=\exp\left(-\delta\int G(\bx-\br)\rho(\br,t)d\br\right),
\label{probc}
\end{equation}
where $\delta$ is a parameter that modulates the strength of the competition,
and $G(\bx)$ is a positive kernel function that introduces a finite range of influence.
This model is related to earlier models of pattern formation in arid systems \cite{Lefever}, and subsequent works \cite{lefeverJTB2009, Lefever2012},
but it differs from standard kernel-based models in that the kernel function accounts for the interaction neighborhood, and not
for the type of interaction with the distance. Note also that the nonlocal term enters nonlinearly in the equation.

Following \cite{justin}, $P_F$ is assumed to be a saturating function of grass biomass, $1-\rho(\bx.t)$, similar to the
implementation of fire of Jeltsch {\it et al.} in \cite{jeltsch}
\begin{equation}
 P_{F}=\frac{\sigma}{\sigma+1-\rho(\bx,t)},
\label{probfire}
\end{equation}
where $\sigma$ governs the resistance to fire, so $\sigma=0$ means no resistance to fires. 
Notice how our model is close to the one in \cite{justin} through the definitions of $P_{C}$ and $P_{F}$, although
we consider the probability of surviving a fire depending on the local density of trees, and in \cite{justin} it
depends on the global density.
The deterministic differential equation that considers tree-tree competition
and fire for the spatial tree density is
\begin{equation}\label{sav1}
 \frac{\partial \rho(\bx, t)}{\partial t}= b_{eff}(\rho)
\rho(\bx, t)\left(1 -\rho(\bx, t)\right)-\alpha\rho(\bx, t),
\end{equation}
where
\begin{equation}
 b_{eff}(\rho)=\frac{b{\rm e}^{-\delta\int G(\bx-\br)\rho(\br,t)d\br}\sigma}{\sigma+1-\rho(\bx,t)}.
\end{equation}

Thus, we have a logistic-type equation with an effective growth rate that depends
nonlocally on the density itself, and which is a combination of long-range competition and local facilitation mechanisms (fire).
The probability of surviving a fire is higher when the local density of trees is higher, as can be seen from the definition
in equation (\ref{probfire}).

\begin{figure}
\centering
\includegraphics[width=0.4\textwidth]{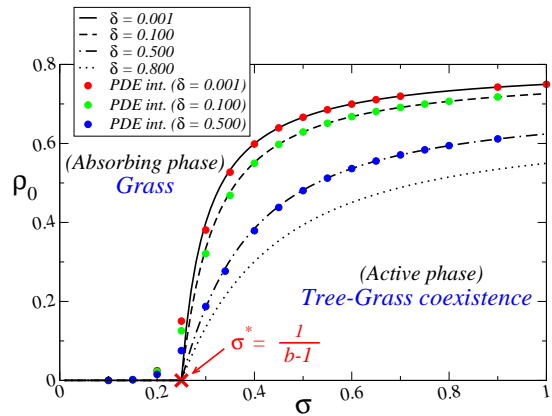}
\caption{Grass-coexistence phase transition. Stationary tree density, $\rho_{0}$, as a function
of the resistance to fires parameter, $\sigma$. The lines come from the mean field solution, Eq.~(\ref{steq}), and the dots
from the numerical integration of Eq.~(\ref{sav1}) over a square region of $1~ha$. We have chosen $\alpha=1$, and $b=5$.
In the case of the spatial model, $\rho_{0}$ involves an average of the density of trees over
the studied patch of savanna.}
\label{transition}
\end{figure}

In figure \ref{transition} we show  numerical solutions for the mean field
equation (\ref{eq:mf}) (lines) and the spatially explicit model (equation \ref{sav1}) (dots) in the stationary state
$(t\rightarrow\infty)$ using different values of the competition. We have used a top-hat function
as the competition kernel, $G(\bx)$ (See section \ref{LSA} for more details on the kernel choice).
We observe a very good agreement of both descriptions which becomes worse when we get closer to the critical point $\sigma^{*}$,
where the model presents a phase transition from a tree-grass coexistence to a grassland state.
This disagreement appears because while the mean field equation describes an infinite system, 
the Eq.~(\ref{sav1}) description forces us to choose a size for the system. 

The model reproduces the long-term coexistence between grass and trees that is characteristic of savannas.
To explore this coexistence, we study the long-time behavior of the system and 
analyze the homogeneous stationary solutions of Eq.~(\ref{sav1}), which has two fixed points.
The first one is the absorbing state representing the absence of trees,
$\rho_{0}=0$, and the other can be obtained, in the general case, by numerically solving  
\begin{equation}\label{steq}
 b_{eff}(\rho_{0})(1-\rho_{0})-\alpha=0.
\end{equation}
In the regime where $\rho_{0}$
is small (near the critical point), if competition intensity, $\delta$, is also small,
it is possible to obtain an analytical expression
for the critical value of the probability of surviving a fire, $\sigma^{*}$, 
\begin{equation}
\sigma^{*}=\frac{\alpha}{b-\alpha}.
\end{equation}
Outside of the limit where $\delta\ll 1$, we can solve Eq.~(\ref{steq}) numerically 
in $\rho_{0}$ to show that the critical value of the fire resistance parameter, $\sigma^{*}$, does not depend on 
competition. A steady state with trees is stable for higher fire survival probability 
(Fig.~\ref{transition}).

The model, then, shows a transition from a state where grass is the only form of vegetation to
another state where trees and grass coexist at $\sigma^{*}$. 
In what follows, we fix $\alpha=1$, so we choose our temporal scale
in such a way that time is measured in units of $\alpha$. This choice does not qualitatively affect our results.

\subsection{Linear stability analysis}
\label{LSA}

The spatial patterns appearing in the nonlocal savanna model can be studied by 
performing a linear stability analysis \cite{cross} of the stationary
homogeneous solutions of equation (\ref{sav1}),
$\rho_{0}=\rho_0 (\sigma,\delta)$. 
The stability analysis is performed by considering
small harmonic perturbations around $\rho_0$,
 $\rho(\bx,t)=\rho_{0}+\epsilon{\rm e}^{\lambda t-i\bk\cdot \bx}$, $\epsilon\ll 1$.
After some calculations (\ref{appA}),
one arrives at the dispersion relation
\begin{eqnarray}\label{reldisper}
& \lambda(k; \sigma, \delta)=b_{eff}(\rho_{0})\frac{1+\sigma(1-2\rho_{0})}{\sigma-\rho_{0}+1}&\nonumber \\
&-b_{eff}(\rho_{0})\frac{\rho_{0}\left[2-\rho_{0}+\delta {\hat G}(k)(\rho_{0}-1)(\rho_{0}-1-\sigma)\right]}{(\sigma-\rho_{0}+1)}-1,&
\end{eqnarray}
where ${\hat G}(k)$, $k =|\bk |$, is the Fourier transform of the kernel,
\begin{equation}\label{foudef}
 {\hat G}(\bk)=\int G(\bx){\rm e}^{-i\bk\cdot \bx}d\bx.
\end{equation}

The critical values of the parameters of the transition to pattern,
$\delta_c$ and $\sigma_c$, and the fastest
growing wavenumber $k_c$ 
are obtained from the simultaneous solution of
\begin{eqnarray}
\lambda (k_c; \sigma_c, \delta_c )&=&0, \label{eq:first}\\
\left(\frac{\partial \lambda}{\partial k}\right)_{k_c; \sigma_c, \delta_c }&=&0. \label{eq:sec}
\end{eqnarray}
Note that $k_c$ represents the most unstable mode of the system, which means that it grows faster
than the others and eventually dominates the state of the system. Therefore,
it determines the length scale of the spatial pattern.
These two equations yield the values of the parameters $\delta$ and $\sigma$ at 
which the maximum of the curve $\lambda (k)$, right at $k_c$, starts becoming positive. This signals
the formation of patterns in the solutions of Eq.~(\ref{sav1}). As
Eq.~(\ref{eq:sec}) is explicitly written as
\begin{equation}\label{derlam}
 \lambda'(k_{c})=b_{eff}(\rho_{0})\delta\rho_{0}{\hat G}'(k_{c})(\rho_{0}-1),
\end{equation}
the most unstable wavenumber $k_c$ can be obtained by 
evaluating the zeros of the derivative of the Fourier transform of the kernel.

Equation~(\ref{reldisper}) shows that competition, through the kernel function,
fully determines the formation of patterns in the system. The local facilitation 
appears in  $b_{eff}(\rho_{0})$ and it is not relevant in the formation of spatial structures.
If the Fourier transform
of $G$ never takes positive values, then $\lambda (k; \sigma, \delta)$ is always
negative and only the homogeneous solution is stable. However, when $\hat G$ can take
negative solutions then patterns may appear in the system. What does this mean in biological
terms? Imagine that we have a family of kernels described by a parameter $p$: 
$G(\bx )= \exp(-|(\bx )/R |^p)$ ($R$ gives the range of competition).
The kernels are more peaked around $\bx =0$ for $p<2$ and more box-like when $p>2$. It turns out that this family
of functions has non-negative Fourier transform for $0 \leq p <2$, so 
that no patterns appear in this case. A lengthy discussion of this property in the context
of competition of species can be found in \cite{pigo}.
Thus, the shape of the competition kernel dictates whether or not patterns will appear in the system. If pattern formation is possible, then the values of the fire and competition parameters
govern the type of solution (see below).

Our central result for nonlocal competition is that, contrary to conventional wisdom, it
can, in the limit of infinitesimally short (purely local) facilitation, promote the clustering of trees. 
Whether or not this occurs depends entirely on the shape 
of the competition kernel. For large $p$ we have a box-like shape, and in these
cases trees compete strongly with other trees, roughly within a distance 
$R$ from their position. The mechanism behind this counterintuitive result is that trees farther than $R$ away from a resident tree area are not able 
to {\it invade} the zone defined by the radius R around the established tree (their seeds do not establish there), so that
an exclusion zone develops around it. For smaller $p$ there is less competition
and the exclusion zones disappear.  

For a more detailed analysis, one must choose an explicit form for the kernel function.
Our choice is determined by the original $P_C$ taken in \cite{justin}, so
that it decays exponentially with the number of trees in a neighborhood of radius $R$ around a
given tree. Thus, for $G$ we take the step function (limit $p\rightarrow\infty$)
\begin{eqnarray}\label{kerneldd}
G(|\br|)= \left\{ \begin{array}{lcc}
             1 &   if  & |\br| \leq R \\
             \\ 0 &  if & |\br| > R. \\
             \end{array}
   \right.
\end{eqnarray}
As noticed before, the idea behind the nonlocal competition is to capture
the effect of the long roots of a tree. The kernel function defines the area of influence
of the roots, and it can be modeled at first order with the constant function of equation (\ref{kerneldd}).
Thus the parameter R, which fixes the {\it nonlocal} interaction scale, must be of the order of the length of the roots \cite{borgogno}.
Since the roots are the responsible for the adsorption of
resources (water and soil nutrients), a strong long-range competition term
implies strong resource depletion.
For this kernel the Fourier transform is \cite{lopezhdez}
${\hat G}(k)=2\pi R^{2} J_{1}(kR)/kR$ 
 and its derivative is
${\hat G}'(k)=-2\pi R^{2} J_{2}(kR)/k$, 
where $ k\equiv |\bk|$, and $J_{i}$ is the $i^{th}$-order Bessel function.
Since ${\hat G}(k)$ can take positive and negative values, pattern solutions
may arise in the system, that will in turn depend on the values of
$\delta$ and $\sigma$.
The most unstable mode is numerically obtained as the first zero of $\lambda'(k)$, Eq.~(\ref{derlam}), which means the first zero
of the Bessel function $J_{2}(kR)$. This value only depends on $R$, being independent of
the resistance to fires and competition, and it is $k_c=5.136/R$.
Because a pattern of $n$ cells is characterized by a wavenumber 
$k_c=2\pi n /L$, where $L$ is the system size,
the typical distance between clusters,
$d_t=L/n$, using the definition of the critical wavenumber is given by $d_t \approx 1.22 R$. In other words, it is approximately
the range of interaction $R$. This result is also independent
of the other parameters of the system.  

Since we are interested in the effect of competition and fire on the distribution of savanna trees,
we will try to fix all the parameters but $\sigma$ and $\delta$. We will explore the effect of different values of these parameters
on the results.
First, we have chosen (as in \cite{justin}) the death rate $\alpha=1$, and solving
Eq.~(\ref{steq}) we will roughly estimate the birth rate, $b$. We will work in the limit of intermediate to high mean annual
precipitation, so water is non-limiting and thus we
can neglect the effects of competition ($\delta=0$). At this intermediate to high mean annual precipitation the empirically 
observed upper limit of savanna tree cover is approximately
$\rho_{0}=0.8$ \cite{sankaran2005, bucini}. To reach this upper limit in the tree cover, disturbances must also be absent, implying
no fire ($\sigma\rightarrow\infty$). In this limit, the mean field equation (\ref{eq:mf}) is quantitatively accurate, as it is shown in Figure \ref{transition},
and the stationary mean field solution of the model depends only on the birth rate
\begin{equation}
\rho_{0} (\sigma \to \infty) =\frac{b-1}{b}.
\end{equation}
It can be solved for $b$ for a fixed $\rho_{0}=0.8$, and it yields $b=5$ \cite{justin}. In the
following we just consider the dependence of our results on $\delta$
and $\sigma$. In particular, $\rho_0=\rho_0 (\sigma, \delta)$.

\begin{figure}
\centering
\includegraphics[width=0.4\textwidth]{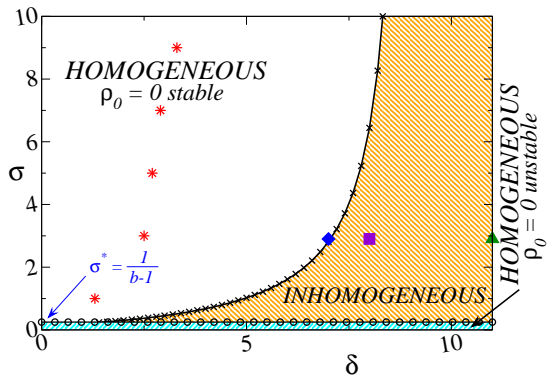}
\caption{Phase diagram of the mean field equation ~(\ref{sav1}) for $b=5.0$, $\alpha=1.0$, and a step kernel.
The absorbing-active transition is shown at $\sigma^{*}$ with circles (o). The homogeneous-pattern transition (Eq.~(\ref{criline})) is
indicated with crosses (x). The diamond, the square, and the up-triangle
show the value of the parameters $\sigma$ and $\delta$ 
taken in Figures~\ref{patterns}(a)-(c)
respectively. The stars point out the transition to
inhomogeneous solutions in the stochastic model as described in Section \ref{stochastic}, with $\Gamma=0.2$.}
\label{phspace}
\end{figure}

The phase diagram of the model, computed
numerically, is shown in Fig.~\ref{phspace},
where we plot 
the spatial character of the steady solution (homogeneous or inhomogeneous) as a function of $\delta$ and $\sigma$.
Note that increasing competition enhances the inhomogeneous or pattern solution.
 This is because, as we are now in the case of a kernel giving rise to clusters, increasing
$\delta$ makes it more difficult to enter the exclusion zones in-between the clusters. 
For very strong competition (high, unrealistic, $\delta$), fire has no influence on the pattern. 

The critical line separating these two solutions (pattern and homogeneous)can be analytically computed as a function of
the parameters $\delta$, $\sigma$, $\rho_0$ and  $\hat{G}(k_{c})$ (see Eq.~(\ref{criline}) in \ref{appB}).
In Figure \ref{phspace} we have plotted (with crosses) this critical line separating homogeneous and pattern
solutions for the step kernel. Note that the stationary density of trees, $\rho_0$, must be computed
numerically from Eq.~(\ref{steq}).

With $b=5$, in the absence of fire ($\sigma\rightarrow\infty$), and
for weak competition, we can take the limits $\delta\rightarrow 0$ and $\sigma\rightarrow\infty$ of 
the dispersion relation Eq.~(\ref{reldisper}), leading to
\begin{equation}\label{limitceroinfinity}
\lambda(k;\delta\rightarrow 0, \sigma\rightarrow\infty)=4-10\rho_{0}.
\end{equation}
In Fig.~\ref{transition}, for large $\sigma$,  it can be seen that typically $\rho_{0}> 0.4$, so Eq.~(\ref{limitceroinfinity}) becomes negative.
This result means that in this limit, trees are uniformly distributed in the system as there is no competition, 
and space does not play a relevant role in the
establishment of new trees. Such situation could be interpreted as favorable to forest
leading to a fairly homogeneous density of trees.
This result agrees with the phase plane plotted in Figure \ref{phspace}. 
In biological terms, there are no exclusion zones in the system because there is no competition.

\subsection{Numerical simulations}
\label{sec:NS}

\begin{figure}
\centering
\includegraphics[width=0.48\textwidth]{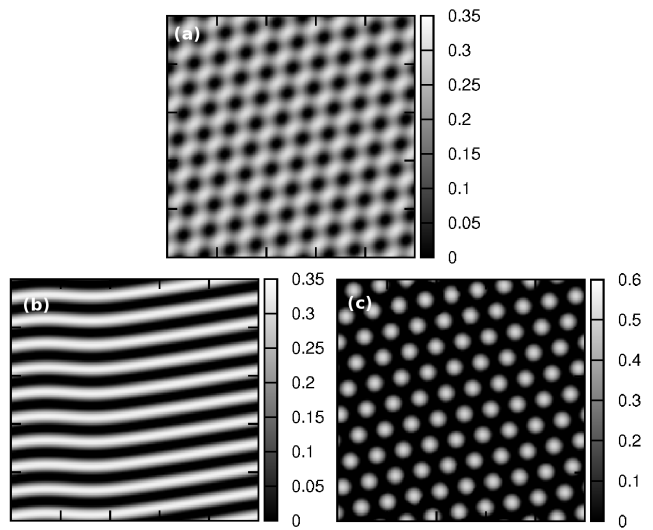}
\caption{(a) Grass spots ($\delta=7.0$), (b) striped grass vs. tree ($\delta=8.0$), and (c) tree spots ($\delta=11.0$) patterns in 
the deterministic model in a square patch of savanna of $1~ha$. $\sigma=2.9$, $R=8.0~m$, $b=5.0$ and $\alpha=1.0$ in all the plots.}
\label{patterns}
\end{figure}

The previous analysis provides information, depending on the competition and fire parameters,
about when the solution is spatially homogenous and when trees arrange in clusters.
However, the different shapes of the patterns have to 
be studied via numerical simulations \cite{dodoricolibro} of the whole equation
of the model. 
We have taken a finite square region of savanna with an area of $1$ ha., allowed competition to occur 
in a circular area of radius $R=8 \ m$, and employed periodic boundary conditions and a finite differences algorithm
to obtain the numerical solution. Similarly to what has been observed in studies of semiarid water limited systems
\cite{dodorico,rietkirkAmNat2002}, different structures, including gaps, stripes, and tree spots,
are obtained in the stationary state as we increase the strength of competition
for a fixed value of the fire parameter or, on the other hand, as we 
decrease the resistance to fires for a given competition intensity. 
In both equivalent cases, we observe this 
spectrum of patterns as far as we go to a more dry state of the system, where resources
(mainly water) are more limited (see Figs.~\ref{patterns}(a)-\ref{patterns}(c)) and competition is consequently stronger. 
This same sequence of appearance of patterns has been already observed in the presence of different short-range
facilitation mechanisms 
\cite{LejeuneJVS1999, rietkirkAmNat2002}. 
It indicates that, when  $\delta$ is increased (i.e. the probability of surviving competition is decreased), new trees cannot establish
in the exclusion areas so clustering is enhanced.

On the other hand, in the case of fire-prones savannas, previous works had only shown either tree spot \cite{LejeunePRE2002} or 
grass spots \cite{dodofire}. Therefore, at some values of the parameter space (see Fig. \ref{patterns}b), 
the patterns in our deterministic approach are not observed
in mesic savannas, and should correspond to semiarid systems. 
However, we will show in the following sections that under the parameter constraints of a mesic savanna, 
and considering the stochastic nature of the tree growth dynamics in the system
(i.e. demographic noise), our model shows realistic spatial structures.

\begin{figure}
\begin{center} 
\includegraphics[width=0.33\textwidth]{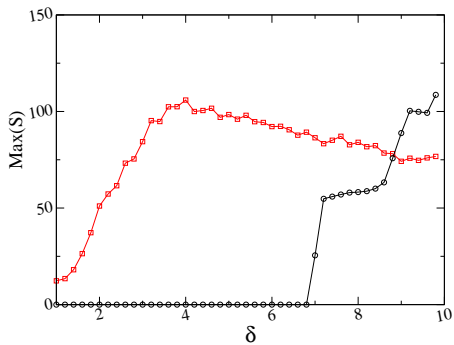}
\caption{Maximum of the structure function for different values of the competition
parameter $\delta$ at long times. The fire parameter is fixed at $\sigma=2.9$. Black
circles refers to the deterministic model and red squares to
the stochastic model, $\Gamma=0.20$.}
\label{maxstructure}
\end{center}
\end{figure}

\begin{figure}
\begin{center}
\includegraphics[width=0.33\textwidth]{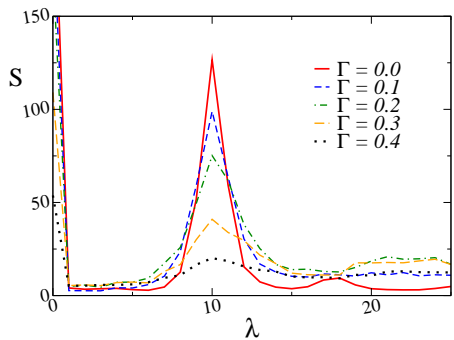}
\caption{Numerical computation of the structure function defined in equation (\ref{strucfunc}) for different values
of the demographic noise intensity. $\delta=9.8$, $\sigma=2.9$, $R=8~m$, $\alpha=1.0$, $b=5.0$.}
\label{strucboth}
\end{center}
\end{figure}

A much more quantitative analysis of the periodicity in the patterns can be performed
via the structure function. This will be helpful to check the previous results and, especially,
for the analysis of the data of the stochastic model of the next section, for which we will not 
present analytical results.
The structure function is defined as the modulus of the spatial
Fourier transform of the density of trees in the stationary state,
\begin{equation}\label{strucfunc}
S(k)=\left\langle \left| \int d\bx{\rm e}^{i\bk \cdot \bx}\rho(\bx,t \to \infty ) \right| \right\rangle,
\end{equation}
where the average is a spherical average over the wavevectors with modulus $k$. 
The structure function is helpful to study spatial periodicities in the system, similar to the power spectrum of a temporal signal. 
Its maximum
identifies dominant periodicities, which in our case are the distances between tree clusters. 
Note that the geometry of the different patterns cannot be uncovered with the structure function, since it
involves a spherical average. In Fig.~\ref{maxstructure}, we show the transition to patterns using the
maximum of the structure function as a function of the competition parameter. A peak appears when 
there are spatial structures in the system, so $Max[S(k)]\neq0$. However, we do not have information about the values where the shapes of the patterns change.
Taking $R=8~m$, the peak is always at $\lambda_c=10 m$ for our deterministic savanna model, independently of the competition
and fire resistance parameters, provided that they take values that ensure the emergence of patterns in the system
(see the line labeled by $\Gamma=0$ in Fig.~\ref{strucboth}; 
for the definition of $\Gamma$ see next section).
This result is in good agreement with the theoretical result provided for the wavelength by the
linear stability analysis $\lambda=2\pi/k_{max}=9.78~m$, which is also independent of competition and resistance to fires.

\section{Stochastic model}\label{stochastic}

The perfectly periodic patterns emerging in Fig.~\ref{patterns} from the deterministic model seem
to be far from the disordered ones usually observed in aerial photographs of mesic savannas and shown by
individual based models \cite{justin, jeltsch1999detecting, barot1999demography, caylor2003tree}. 
We have so far described a savanna system in terms of the density of trees with a deterministic
dynamics. The interpretation of the field $\rho(\mathbf{x},t)$ is the density of tree (active) 
sites in a small volume, $V$. If we think of trees as reacting particles
which are born and die probabilistically, then to provide a reasonable description of the underlying individual-based
birth and death dynamics,
we have to add a noise term to the standard deterministic equation. It  
will take into account the {\it intrinsic} stochasticity present at the individual level in the system.

If we take a small volume, $V$, the number of reactions taking place is proportional to the number of particles therein, $N$, with 
small deviations. If $N$ is large enough, the central limit theorem applies to the sum of $N$ independent random
variables and predicts that the amplitude of the deviation is of the order of $\sqrt{N}\propto\sqrt{\rho(\mathbf{x},t)}$ \cite{gardiner}.
This stochasticity referred to as demographic noise. The macroscopic equation is now stochastic,
\begin{eqnarray}\label{savsto}
   \frac{\partial \rho(\mathbf{x}, t)}{\partial t}&=&b_{eff}(\rho)[\rho(\mathbf{x}, t)-\rho^{2}(\mathbf{x}, t)]- \nonumber \\
&-&\alpha\rho(\mathbf{x}, t)+\Gamma\sqrt{\rho(\mathbf{x}, t)}\eta(\mathbf{x},t),
\end{eqnarray}
where $\Gamma \propto \sqrt{b_{eff}}$ (but we take it as a constant, \cite{dickman}) modulates the intensity of $\eta(\mathbf{x},t)$,
 a Gaussian white noise term with zero mean and correlations given by Dirac delta distributions
\begin{equation}
 <\eta(\mathbf{x},t)\eta(\mathbf{x'},t')>=\delta(\mathbf{x}-\mathbf{x'})\delta(t-t').
\end{equation}
The complete description of the dynamics in Eq.(\ref{savsto})
should have the potential to describe more realistic patterns.

\begin{figure}
\begin{center} 
\includegraphics[width=0.33\textwidth]{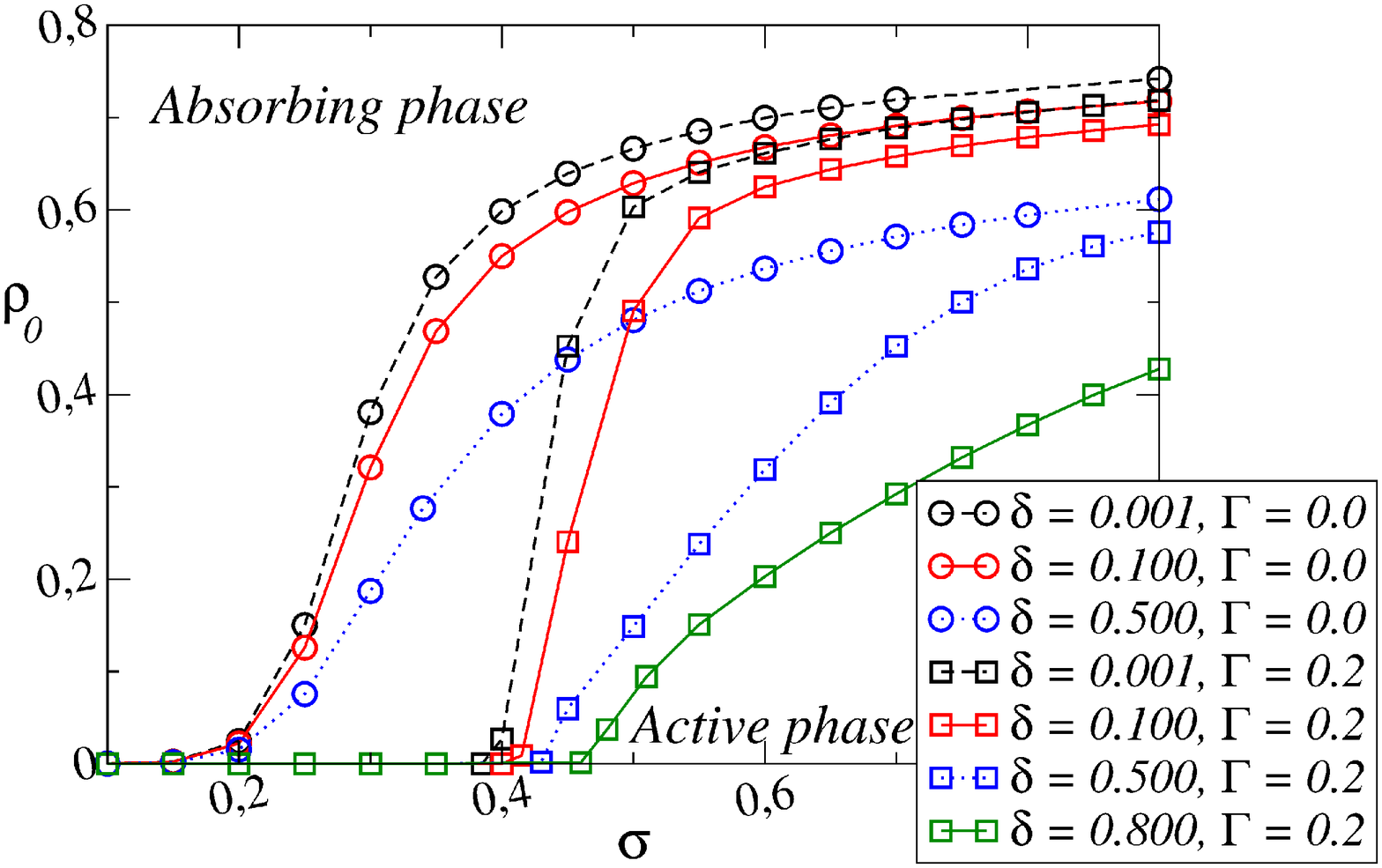}
\caption{Active-absorbing phase transition in the deterministic (Circles) and the stochastic
model (Squares). In the later case, we integrate the Eq.~(\ref{savsto}) with $\Gamma=0.2$ and
average the density of trees in the steady state.}
\label{noisy-transition}
\end{center}
\end{figure}

We first investigate the effect of demographic noise  on the persistence of trees in the system.
We show in (Fig. \ref{noisy-transition}) that the critical point, $\sigma^{*}$, depends on the value of the
competition parameter $\delta$. This effect is rather small, so that when $\delta$ increases the transition to the
grassland state appears only for a slightly larger $\sigma$ (i.e, less frequent fire). The reason seems to be that fire
frequency and intensity depend on grass biomass. Seasonally wet savannas support much more grass biomass that serves as 
fuel for fires during the dry season \cite{dodo2, hanan2008}. Dry savannas have much lower grass biomass, so they do not burn as often or as intensely.
The shift of the critical value of $\sigma$ when competition is stronger
is consistent with the one showed in \cite{justin}, as can be seen comparing Figure 2 in \cite{justin} 
with Figure \ref{noisy-transition} here. Besides, the values obtained for $\sigma^{*}$ are larger when we consider the demographic stochasticity \cite{stanley}
neglected in the deterministic field approach.

\begin{figure}
\centering
\includegraphics[width=0.48\textwidth]{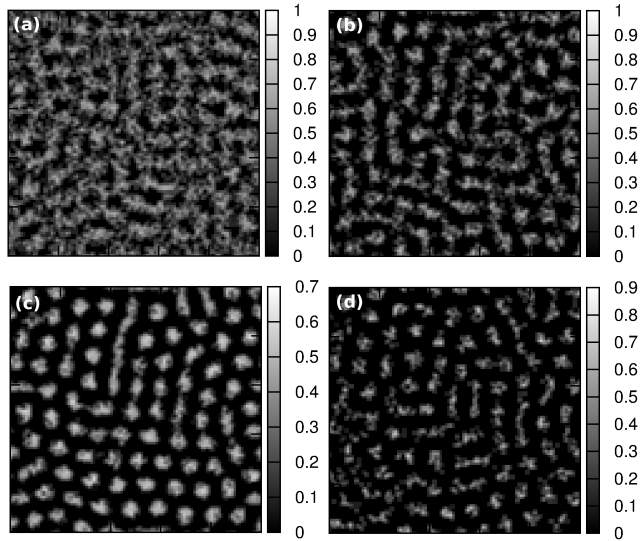}
\caption{Patterns of the stochastic model in a square patch of savanna of $1~ha$. $\sigma=2.9$, $R=8.0~m$, $b=5.0$ and $\alpha=1.0$ in all the plots.
 (a) $\Gamma=0.2$, $\delta=3.0$. (b) $\Gamma=0.2$, $\delta=5.0$. (c) $\Gamma=0.1$, $\delta=10.0$. (b) $\Gamma=0.2$, $\delta=10.0$.}
\label{stopatterns}
\end{figure}

We explore numerically the stochastic savanna model using an algorithm developed in
\cite{dickman} (See \ref{sec:numerical}).
Note that the noise makes the transition to pattern smoother so the change
from homogeneous to inhomogeneous spatial patterns is not as clear as it is in the limit
where the demographic noise vanishes (See Fig.~\ref{maxstructure}). The presence of
demographic noise in the model, as shown in Fig.~\ref{phspace} (red stars),
also decreases the value of the competition strength at which
patterns appear in the system, as has been observed in other systems. Mathematically, these new patterns appear since 
demographic noise maintains Fourier modes of the solution which, due to the value of the parameters,
would decay in a deterministic approach \cite{butler}. Biologically, exclusion zones are promoted by demographic noise, since it does not 
affect regions where there are not trees. On the other hand in vegetated areas fluctuations may enhance tree density, leading to stronger competition.
The presence of demographic noise in the model allows the existence of patterns under more humid conditions.
This result is highly relevant for mesic savannas, as we expect competition to be of low to intermediate strength in such systems.
We show two examples of these irregular patterns in Fig.~\ref{stopatterns}(a) and Fig.~\ref{stopatterns}(b). Unrealistic stripe-like patterns
no longer appear in the stochastic model. 

We have studied the dynamics of the system for some values of the fire and 
competition parameters. Demographic noise influences the spatial 
structures shown by the model. The deterministic approach shows a full spectrum of
patterns which are not visually realistic for mesic savannas (but for arid systems). The role of
the noise is to transform this spectrum of regular, unrealistic patterns into more irregular ones 
(Figures \ref{stopatterns}(a)-\ref{stopatterns}(d))
that remind the observed in aerial photographs of real mesic savannas. On the other hand, these patterns
are statistically equivalent to the deterministic ones, as it is shown with the structure function in Fig. \ref{strucboth}.
The dominant 
scale in the solution is given by the interaction radio, $R$, and it is independent of the 
amplitude of the noise (see the structure function in Figure \ref{strucboth},
peaked around $\lambda=10~m$ independently of the noise). Besides, over a certain treshold in the amplitude, demographic
noise destroys the population of trees. Therefore, the model presents
an active-absorbing transition with the noise strength, $\Gamma$, being the control parameter. 

\section{Discussion}
\label{sec:discussion}

Understanding the mechanisms that produce spatial patterns in savanna tree populations has long been an area of interest among savanna ecologists 
\cite{skarpe1991spatial, jeltsch1999detecting, barot1999demography, caylor2003tree, scanlon2007positive}.
A key step in such an analysis is defining the most parsimonious combination of mechanisms that will produce the pattern in question. 
In this paper the combination of long-range competition for resources and the facilitation induced by fire  
are considered the responsible of the spatial structures, in the line of studies of vegetation pattern formation
in arid systems, where also a combination of long-range inhibition and short-range facilitation is introduced \cite{klausmeier, Lefever, rietkirkAmNat2002, vonHardenberg}.
The main difference is that the facilitation provided by the protection effect of adult trees against fires in our savanna model takes the short-range facilitation to 
its infinitesimally short limit (i.e, local limit). Under this assumption we have studied the conditions under which our model could account for patterns.
We have shown that nonlocal competition combined with local facilitation induces the full range of
observed spatial patterns, provided the competition term enters nonlinearly 
in the equation for the density of trees,
and that competition is strong enough.

The key technical requirement for this effect to occur is that the competition kernel must be an almost constant function in a given competition region, and decay
abruptly out of the region. We verify this condition working with supergaussian kernel functions.
In practice, this means that competition kernels whose Fourier transform takes 
negative values for some wavenumber values, will lead to competition driven clustering.

The other mechanism we have considered for a minimalistic but realistic
savanna model, fire, has been shown to be relevant
for the coexistence of trees and grass and for the shape of the patterns.
However, competition is the main ingredient allowing pattern solutions to exist in the model. If the shape of the kernel
allows these types of solutions, then the specific values of fire and competition
parameters determine the kind of spatial structure that develops.
It is also worth mentioning that one can observe the full spectrum of patterns in the limit where fires vanish 
($\sigma\rightarrow\infty$), so there is no facilitation at all, provided competition is strong enough. However,
when there is no competition, $\delta=0$, no patterns develop regardless of the value of the fire term.
Therefore, we conclude that the nonlocal competition term is responsible for the emergence of clustered distributions 
of trees in the model, with the fire term playing a relevant role only to fix the 
value of the competition parameter at which patterns appear. In other words, for a given competition strength, 
patterns appear more readily when fire is combined with competition. A similar mechanism of competitive interactions
between species has been shown to give rise to clusters of species in the context of classical ecological niche theory. Scheffer and 
van Nes \cite{scheffer2006} showed that species distribution in niche space was clustered, and Pigolotti et al. \cite{pigo} showed that 
this arises as an instability of the nonlocal nonlinear equation describing the competition of species.

Long-distance competition for resources in combination with the local facilitation due to the protection effect
of adult trees in the establishment of juvenile ones can explain the emergence of realistic structures of trees in mesic savannas. 
In these environmental conditions, competition is limited, so we should restrict to small to intermediate values of the parameter $\delta$, 
and the effect of fires is also worth to be taken into account. However, these two ingredients give a full range of patterns
observed in vegetated systems, but not in the particular case of savannas. It is necessary to consider the role of demographic noise, 
which is present in the system through the stochastic nature of the birth and death processes of individual trees. In this complete
framework our model shows irregular patterns of trees similar to the observed in real savannas.

The other important feature of savannas, the characteristic long-time coexistence of trees and grass is well
captured with our model (Figures \ref{transition} and \ref{noisy-transition}). Besides, the presence of 
demographic noise, as it is shown in Figure \ref{noisy-transition}, makes our approach much 
more realistic, since the persistence of trees in the face of fires is related to the water in the system.
On the other hand, demographic stochasticity causes tree extinction at lower fire frequencies (larger $\sigma$) than
in the deterministic case. This is because random fluctuations in tree density are of sufficient
magnitude that this can hit zero even if the deterministic stationary 
tree density (for a given fire frequency) is greater than zero. This effect vanishes if
we increase the system size. The demographic noise is proportional to the density of trees (proportional
to $(L_{x}\times L_{y})^{-1}$), so fluctuations are smaller if we study bigger patches of savannas.
As usually happens in the study of critical phenomena in Statistical Mechanics, the extinction times due to 
demographic noise increase exponentially with the size of the system for those intensities of competition
and fire that allow the presence of trees in the stationary state. Over the critical line, this time will follow
a power law scaling, and a logarithmic one when the stationary state of the deterministic model is already absorbing 
(without trees) \cite{marro}.

\section{Summary} \label{summary}

We have shown the formation of patterns in a minimal savanna model, that 
considers the combination of long-range competition and local facilitation mechanisms
as well as the transition from  
trees-grass coexistence to a grass only state. 
 
The salient feature of the model is that it only considers
nonlocal (and nonlinear) competition through a kernel function which defines the length
of the interaction, while the facilitation is considered to have an infinitesimally short influence range.
Our model thus differs from standard kernel-based savanna models that feature both
short-range facilitation and long-range competition. The same sequence of spatial patterns
appears in both approaches, confirming Rietkerk and van de Koppel's \cite{rietkirkRev2008} suggestion
that short-range facilitation does not induce spatial pattern formation by itself, and long-distance competition
is also needed. It also suggests that long-range competition could be not only a necessary, but also a 
sufficient condition to the appearance of spatial structures of trees.

Inspired by \cite{justin}, we have proposed a nonlocal deterministic macroscopic equation 
for the evolution of the local density of trees where fire and
tree-tree competition are the dominant mechanisms.
If the kernel function falls off with distance very quickly (the Fourier
transform is always positive) the system only has homogenous solutions. 
In the opposite case, patterns may appear depending on the value of the parameters ($\delta$ and $\sigma$),
and in a sequence similar to the spatial
structures appearing in standard kernel-based models. Under less favorable environmental
conditions, trees tend to arrange in more robust structures to survive 
(Fig.~\ref{patterns}(d)). Biologically, trees are lumped
in dense groups, separated by empty regions. 
Entrance of new trees in these {\it exclusion zones} is impossible
due to the intense competition they experience there.

A great strength of our approach is that our deterministic analysis is formal, 
and we have shown the different spatial distributions of the
trees that occur as competition becomes more intense, concluding that self organization of trees is a good
mechanism to promote tree survival under adverse conditions \cite{rietkirkAmNat2002}. 
Trees tend to cluster in the high competition (low resources) limit (Fig.~\ref{patterns}(d)), due to the 
formation of exclusion zones caused by non-local competition, and not as a result of facilitation.
However, because we are dealing with a deterministic model, the patterns
are too regular and the transition between the grass-only and a tree-populated
states is independent of tree competition. We therefore considered stochasticity
coming from the stochastic nature of individual birth and death events,
to provide a more realistic description of savanna dynamics. Calabrese et al. \cite{justin} also noted that savanna-to-grassland
transition was independent of competition intensity in the mean field approach, but not when demographic noise was included. In
the present model, both the grassland to savanna transition 
and the spatial structures that develop are influenced by demographic stochasticity. In the case of spatial 
structures, demographic noise is specially relevant, since it turns much of the unrealistic patterns of the deterministic 
model into more realistic ones, that remind the observed in real savannas. It also allows the existence of periodic arrangements 
of trees in more humid systems, which means environmental conditions closer to mesic savannas.

We have quantified the characteristic spacing of spatial patterns
through the structure function. The irregular patterns produced by the stochastic model still
have a dominant wavelength whose value is the same as in the
deterministic model and depends only on the value of the range 
of the interaction, $R$, in the kernel function. The match between the typical spatial scale of the
patterns and the characteristic distance over which nonlocal competition acts indicates
that competition is responsible for the presence of clustered spatial structures.

\section{Acknowledgments}
R.M-G is supported by the JAEPredoc program of CSIC. 
R.M-G. and C.L. acknowledge support 
from MICINN
(Spain) and FEDER (EU) through Grant No. FIS2007-
60327 FISICOS.
We acknowledge Federico V\'azquez and Emilio Hern\'andez-Garc\'ia for their comments and discussion.
We also acknowledge the detailed reading and insightful comments of three anonymous referees which greatly helped
to improve this manuscript.
\appendix
\section{Linear stability analysis}
\label{appA}

This appendix shows the details of the linear stability analysis, in particular how
it is obtained the dispersion relation in Eq.~(\ref{reldisper}). We 
consider the stationary solution $\rho_{0}$ plus a small harmonic perturbation, 
\begin{equation}\label{ansatz}
 \rho(\bx,t)=\rho_{0}+\epsilon {\rm e}^{\lambda t-i\bk\cdot\bx},
\end{equation}
where $\epsilon\ll 1$. Substituting Eq.~(\ref{ansatz}) into the original equation (\ref{sav1}), and
retaining only linear terms in $\epsilon$,we arrive to the relation dispersion

\begin{eqnarray} \label{rd}
&\lambda(k)=bC\sigma(\rho_{0}-\rho_{0}^{2})\left[\frac{1}{(\sigma+1-\rho_{0})^{2}}-\frac{\hat{G}(k)\delta}{\sigma+1-\rho_{0}}\right]& \nonumber \\
&+bC\sigma\frac{1-2\rho_{0}}{\sigma+1-\rho_{0}}-1,& \nonumber \\
\end{eqnarray}
where $\hat{G}(k)$ is the Fourier transform of the kernel, $\hat{G}(k)=\int G(x)\exp(\lambda t-i\bk\cdot\bx)$, and 
$C\equiv\exp\left(-\delta\rho_{0}\right)$, provided that we deal with normalized kernels.
Equation (\ref{rd}) can be written as Eq.~(\ref{reldisper}) using the definition of $b_{eff}(\rho_{0})$.

\section{Expression of the transition to pattern critical line.}
\label{appB}

We show here the analytical expression for the critical line in the transition
from homogeneous to inhomogeneous solutions. Starting from Eq.~(\ref{eq:first})
it is possible to write an expression for the value of the resistance to fires
parameter, $\sigma$, at which the macroscopic equation (\ref{sav1}) starts
showing pattern solutions,
as a function of the competition parameter, $\delta$, and the most unstable mode 
$k_{c}$. Considering the value of the parameters taken in our study, $b=5$ and $\alpha=1$, it is
\begin{eqnarray}\label{criline}
&\sigma_c=\frac{(\rho_{0}-1)[5(\rho_{0}-1)
(\delta \hat{G}(k_{c})\rho_{0}-1)-2{\rm e}^{\delta\pi R^{2}\rho_{0}}]}{10\left[1-2\rho_{0}+
\delta \hat{G}(k_{c})\rho_{0}(1+\rho_{0})-{\rm e}^{\delta\pi R^{2}\rho_{0}}/5\right]}& \nonumber \\
&+ \frac{(\rho_{0}-1)\sqrt{5\left[5(\rho_{0}-1)^{2}
(\delta \hat{G}(k_{c})\rho_{0}-1)^{2}-4{\rm e}^{\delta\pi R^{2}\rho_{0}}\rho_{0}\right]}}{10\left[1-2\rho_{0}+
\delta \hat{G}(k_{c})\rho_{0}(1+\rho_{0})-{\rm e}^{\delta\pi R^{2}\rho_{0}}/5\right]}.&
\end{eqnarray}

This complicated expression must be evaluated numerically together with the 
solution of Eq.~(\ref{steq}) for the stationary density of trees, which is also
a function of the competition and fire parameters. We show the results 
in Figure \ref{phspace}, where the curve, represented with the black crosses, 
fits perfectly with the numerical results from the linear stability analysis. 

\section{Numerical algorithm for the integration of the equation (\ref{savsto}).}
\label{sec:numerical}

The integration of stochastic equations where the noise amplitude depends on the square root of the 
variable, $\rho$, and there are absorbing states (i.e, states where the system stays indefinitely),
has awaken a great interest, specially in the study of critical phenomena (i.e, properties of the system that appear
when it is close to the critical point, often the absorbing state). The 
amplitude of the fluctuations tends to zero there, and thus numerical instabilities may appear. Recently \cite{sto1, sto2}
a very efficient method has been developed, but we have used in this work an older one, presented in \cite{dickman}, since its 
implementation is easier and it gives precise results working far from the transition point. It consists on
discretizing the Langevin equation, taking a step size $\Delta\rho$ in the variable.

To apply the method to equation (\ref{savsto}), first of all we discretize the space. Particularly, we
compute the integral in the exponential term approximating it by a sum of the field evaluated in the nodes of the discrete space
\begin{equation}
 \int\rho(\bx,t)G(\bx-\bx')dx\approx\sum_{i=1}^{N_{x}}\sum_{j=1}^{N_{y}}\rho_{i,j}G_{i,j;i',j'}\Delta x\Delta y.
\end{equation}

Then, we integrate the temporal dependence. The key of the algorithm is to prevent $\rho+\Delta\rho$ to take negative values. From a general equation
\begin{equation}
 \frac{d\rho}{dt}=f(\rho)+\sqrt{\rho}\psi(t),
\end{equation}
where $\psi(t)$ is a gaussian white noise with zero mean and delta correlated, it is
\begin{equation}
 \Delta\rho=f(\rho)\Delta t+\sqrt{\rho}\Delta W,
\end{equation}
where $\Delta W=\sqrt{\Delta t}Y$. $Y$ is a Gaussian number with zero mean and unit variance.
At this point, to prevent $\rho+\Delta\rho$ to take negative values, the author in \cite{dickman}
proposes to dicretize the density setting $\rho=n\rho_{min}$ and to truncate the gaussian distribution from
where $Y$ is obtained simetrically so that
$|Y|\leq Y_{max}$. The negatives values are avoided requiring $Y_{max}\sqrt{\Delta t}\leq\rho_{min}$.
It can be done in many ways but following \cite{dickman} we use
\begin{eqnarray}
 Y_{max}&=&\frac{|\ln \Delta t|}{3}, \nonumber \\
 \rho_{min}&=&\frac{(\ln \Delta t)^{2}\Delta t}{9}.
\end{eqnarray}
Finally, rescaling the equation, we can achieve a discretized version in which positive
and zero-mean noise are ensured at the cost of a ``quantized'' density.

\section{The effect of rainfall: Random switching between death and birth}
\label{sec:rainfall}

One of the key ingredients for the long coexistence
between grass and trees is the largely inhomogeneous
temporal distribution of precipitations over time
\cite{sankaran2005, vazquez, dodoricolibro}. We have studied
 this environmental variability following
the idea in \cite{dodorico2006b}, considering the
switching between unstressed vegetation growth,
given by the first term in (\ref{sav1}), and 
drought-induced vegetation decay, represented with 
the second term in Eq.~(\ref{sav1}). These processes
take place each time step with probability $P$ and $1-P$,
respectively. From now on, we call
\begin{eqnarray}
 f_{b}[\rho(\mathbf{x},t)]&=&b_{eff}(\rho)\left[\rho(\mathbf{x},t)-\rho^{2}(\mathbf{x},t)\right], \nonumber \\
 f_{d}[\rho(\mathbf{x},t)]&=&-\alpha\rho(\mathbf{x},t),
\end{eqnarray}
and
\begin{equation}\label{fpm}
 f_{\pm}[\rho(\mathbf{x},t)]=\frac{1}{2}\left[f_{b}[\rho(\mathbf{x},t)]\pm f_{d}[\rho(\mathbf{x},t)]\right].
\end{equation}

The random dynamics of the system is written in terms of a stochastic partial differential equation,
\begin{equation}\label{dichsde}
 \frac{\partial\rho(\mathbf{x},t)}{\partial t}=f_{+}[\rho(\mathbf{x},t)]+f_{-}[\rho(\mathbf{x},t)]\xi_{dn}(t),
\end{equation}
where $\xi_{dn}(t)$ is a dichotomous noise (DMN), assuming values $+1$ (wet season) 
and $-1$ (dry season) with probability $P$ and $1-P$, respectively.

If the rate of random switching, taken as the inverse of the
integration time step, is relatively fast 
respect to the rate of convergence to equilibrium in each of the 
two states, we can replace the noise term in Eq.~(\ref{dichsde}) 
with its average
value, $<\xi_{dn}(t)>=1-2P$. It is meaningful since the rainfall 
seasons are much shorter than the time needed to reach one of the equlibrium stationary
states of death and birth processes, $\rho(\mathbf{x},t)=0,1$, 
respectively. This substitution leads to a deterministic equation
\begin{equation}\label{dichpde}
 \frac{\partial\rho(\mathbf{x},t)}{\partial t}=f_{+}[\rho(\mathbf{x},t)]+f_{-}[\rho(\mathbf{x},t)](1-2P),
\end{equation}
where we will be able to perform linear stability analysis as
 usual. The new dispersion relation is easily obtained,
\begin{eqnarray}\label{reldisperswitch}
&\lambda(k; \sigma, \delta)=b_{eff}(\rho_{0})P\frac{1+\sigma(1-2\rho_{0})}{\sigma-\rho_{0}+1}-(1-P)&\nonumber \\
&-b_{eff}(\rho_{0})P\frac{\rho_{0}\left[2-\rho_{0}+\delta {\hat G}(k)(\rho_{0}-1)(\rho_{0}-1-\sigma)+2\sigma\right]}{(\sigma-\rho_{0}+1)},&
\end{eqnarray}
which means that the main effect of the dichotomous noise is to 
renormalize the rates $\alpha$ and $b$. The patterns observed
now are the same as the ones in the deterministic case, though
the regions where they emerge change in accordance with this
renormalization. Thus, the effect of stochastic precipitation,
as modeled with this random switching mechanism, is a change of
the parameter values for the different transitions observed in
the deterministic continuum model Eq.~(\ref{sav1}).

According to the value of $P$, an absorbing-active phase transition is 
observed, $P_{c}\approx0.20$. Small values of P, meaning long 
dry season, lead to an absorbing state while increasing the
probability of raining implies the appearence of trees in the system. 
In this latter case, the solution can be either homogeneous or showing
 spatial patterns, depending on fire and competition.

This attempt to model rainfall has not been very succesful 
and does not give a lot of new information. Much effort of
future research should be put on this point, trying to get
much more realistic modelling of external environmental 
variability, according to empirical observations, with long
runs of dry years and rare wet years.
%----------------------------------------------------------------

\end{document}